\documentclass{jaa}
\usepackage[authoryear]{natbib}
\usepackage{graphicx}
\usepackage{url}
\usepackage{xcolor}
\usepackage{array}
\usepackage[colorlinks=true,citecolor=blue]{hyperref} 
\usepackage{booktabs}
\usepackage{fontawesome5}
\definecolor{orcidgreen}{HTML}{A6CE39}

\newcolumntype{C}[1]{>{\centering\arraybackslash}p{#1}}

\newcommand{\sun}     {$_{\odot}$}

\begin{document}\sloppy

\title{On the detectability and parameterisation of binary stars through spectral energy distributions}


\author{Vikrant V. Jadhav\href{https://orcid.org/0000-0002-8672-3300}{\color{orcidgreen}\faOrcid{}}
\textsuperscript{1,*}}
\affilOne{\textsuperscript{1}Helmholtz-Institut für Strahlen- und Kernphysik, Universität Bonn, Nussallee 14-16, D-53115 Bonn, Germany}

\twocolumn[{

\maketitle

\corres{vjadhav@uni-bonn.de}

\msinfo{23 September 2024}{5 December 2024}

\begin{abstract}
This study examines the characterization of binary star systems using Spectral Energy Distributions (SEDs), a technique increasingly essential with the rise of large-scale astronomical surveys. Binaries can emit flux at different regions of the electromagnetic spectrum, making SEDs a valuable tool in identifying and characterising unresolved binary systems. However, fitting multi-component models to SEDs and recovering accurate stellar parameters remains challenging due to nonlinear fitting methods and inherent uncertainties in the data and the spectral models.
In this work, a simplified approach was used to model stars as blackbodies and we tested the accuracy of parameter recovery from SEDs, particularly focusing on secondary stars. We explored a range of primary properties, filter sets and noise models. 
Special attention was given to two case studies: one examining the detection of unresolved binaries using \textit{Gaia} XP spectra, and the other focusing on identifying hotter companions in binary systems using UV-IR SEDs.
Although an analytic prescription for recoverability is not possible, we present a simplified model and the necessary \textsc{python} tools to analyse any potential binary system.
Finally, we propose using blackbody models as a baseline for error estimation in SED fitting, offering a potential method for measuring fitting errors and improving the precision of binary star characterisations.
\end{abstract}

\keywords{binaries: general--- techniques: photometric---Spectral energy distribution}

}]


\doinum{12.3456/s78910-011-012-3}
\artcitid{\#\#\#\#}
\volnum{000}
\year{0000}
\pgrange{1--}
\setcounter{page}{1}
\lp{1}

\section{Introduction}

The majority of the stars in the Galaxy are in binary or higher order systems \citep{Moe2017ApJS..230...15M, Luo2021arXiv210811120L}. Different techniques are used to identify different types of binaries: multi-epoch astrometry to detect astrometric binaries, multi-epoch photometry to detect eclipsing binaries, multi-epoch spectroscopy to detect spectroscopic binaries and multi-band imaging to detect unresolved photometric binaries. With the advent of large-scale surveys, a large fraction of the sky has been covered in multiple wavelengths. Using this vast amount of data to detect and characterise the binaries is important to understand the present-day binary population. 

This work focuses on the detection and characterisation of binaries using spectral energy distributions (SEDs). SEDs are a popular tool to characterise stars \citep{Bayo2008A&A...492..277B, Jadhav2019ApJ...886...13J, Thompson2021AJ....161..160T, Rao2022MNRAS.516.2444R} and galaxies \citep{Conroy2010ApJ...712..833C, Boquien2019A&A...622A.103B} due to the efficiency of using photometric data. Among stellar sources, unequal temperature binary components emit flux in different regions of the electromagnetic spectra, thus SEDs are particularly helpful in identifying such unresolved binaries. Most studies fit the source SEDs with multi-component spectra and recover the best-fit stellar parameters and use either the model grid spacing \citep{Gosnell2015ApJ...814..163G, Lodieu2019A&A...628A..61L} or statistical bootstrapping \citep{Jadhav2019ApJ...886...13J}, or $\chi^2$ based techniques \citep{Nine2023ApJ...944..145N} to estimate errors. However, judging the accuracy of these fits is non-trivial. Furthermore, the uncertainty estimation for the parameters is also complicated due to the non-linear nature of the SED fitting techniques \citep{Andrae2010arXiv1009.2755A, Andrae2010arXiv1012.3754A}. Thus, there is a need to establish the parameter space where the SEDs give accurate parameters and estimate the typical errors associated with the primary-secondary properties.

In this work, we test the recoverability of secondary parameters and their errors using SEDs. In Section~\ref{sec:method} we describe the method to create and fit binary sources assuming the stars were blackbodies. Section~\ref{sec:general_hrd} presents the overall results for the recoverability for primaries and secondaries across a grid in the Hertzsprung–-Russell diagram (HRD). Section~\ref{sec:case_unresolved} and \ref{sec:case_uocs} present the case studies for recovering unresolved main sequence (MS) binaries using \textit{Gaia} XP spectra and hotter companions using UV-IR SEDs, respectively. Section~\ref{sec:case_errors} demonstrates a method to estimate blackbody-based error limits for SED parameters.

\vspace{-1em}
\section{Method} \label{sec:method}

\subsection{Creating Model SEDs}

Stars are not perfect blackbodies thus their spectra are slightly different than a blackbody's. There are multiple model spectra available for different spectral types: brown dwarfs  \citep{Allard2001ApJ...556..357A, Allard2011ASPC..448...91A}, MS and red giants \citep{Allard2003IAUS..211..325A, Coelho2014MNRAS.440.1027C, Castelli2003IAUS..210P.A20C}, white dwarfs (WDs; \citealt{Koester2010MmSAI..81..921K, Levenhagen2017ApJS..231....1L}), Y dwarfs \citep{Morley2014ApJ...787...78M}, subdwarfs \citep{Pacheco2021ApJS..256...41P}.
However, there is no spectral model which covers the complete range in temperature, log($g$) and luminosity homogeneously. Furthermore, there is an overlap between the parameter range covered by various models and there are inherent inaccuracies in the models based on their respective assumptions. Thus, using a combination of different spectral models and maintaining the specific model for a particular star is non-trivial. To simplify the process and to avoid any model-specific issues, we assumed that all stars were blackbodies and their flux was only a function of temperature and radius.
The flux of a star with a distance \textit{D}, temperature \textit{T}, radius \textit{R} ($\equiv$ luminosity, $L=4 \pi R^2 \sigma_{sb}T^4$ ) observed using a filter of pivot wavelength of $\lambda$ can be calculated as follows.
\footnote{Note added after acceptance: The published version assumes the source was observed through an infinitesimally narrow filter. Support for taking filter transmission curves as inputs and convolving the blackbody flux was added during further development. The broader filters flatten the SED and the fitting errors increase by a fraction. Hence, the lower limit on fitting errors based on the published method (infinitesimally narrow filter) remains valid.}
\begin{equation} \label{eq:flux_single}
    F(T, R, D, \lambda) = \left(\frac{2hc^2}{\lambda^5} \frac{1}{exp\left( \frac{hc}{\lambda k T}\right) - 1}\right) \left(\frac{R}{D}\right)^2
\end{equation}
For simplicity, we used $D=10$ pc for the rest of the analysis.
We also simulated observational noise in the data using a simple Gaussian noise. The Gaussian noise with a mean of 0 and standard deviation of $\sigma$ ($\equiv$ fractional errors) is denoted by $\mathcal{N}(0, \sigma^2)$. The $\sigma$ can be a single noise level applicable to all filters or can be unique for individual filters. The noisy flux was defined as follows.

\begin{equation}
    F_{noisy}(T, R, \sigma, \lambda) = F(T, R, \lambda) \times (1+\mathcal{N}(0, \sigma^2))
\end{equation}

To create binaries, we simply combine two blackbodies as follows.
\begin{equation}
\begin{split}
    F_{bin}(T_1, T_2, R_1, R_2, \lambda) &= F(T_1, R_1, \lambda) + F(T_2, R_2, \lambda)\\
    F_{bin,\, noisy}(\sigma) &= F_{bin}\times (1+\mathcal{N}(0, \sigma^2))
\end{split}
\end{equation}

Ultimately, the SED of a star or a binary can be constructed for a given set of filters. In this work, we use 5 types of filter sets.

\begin{enumerate}
    \item \texttt{optimistic\_filters}: 50 filters within the range of 1350--50000 $\AA$ (equidistant in log-space).
    \item \texttt{realistic\_filters}: 15 filters within the range of 1350--50000 $\AA$ (equidistant in log-space).
    \item \texttt{gaia\_xp\_filters}: The wavelength range of \textit{Gaia} XP sampled mean spectrum consisting of 343 filters (336 to 1020 nm with a step of 2 nm).
    \item \texttt{uocs\_filters}: A set of 24 filters from UltraViolet Imaging Telescope (UVIT/\textit{AstroSat} \citealt{Tandon2017AJ....154..128T}), GALEX \citep{Martin2005ApJ...619L...1M}, Sloan Digital Sky Survey (SDSS \citealt{Doi2010AJ....139.1628D}), \textit{Gaia} \citep{Gaia2016A&A...595A...1G}, Two Micron All Sky Survey (2MASS \citealt{Skrutskie2006AJ....131.1163S}) and \textit{Wide-field Infrared Survey Explorer} (\textit{WISE} \citealt{Wright2010AJ....140.1868W}).
    \item \texttt{optical\_IR\_filters}: A set of 15 filters from SDSS, \textit{Gaia}, 2MASS and \textit{WISE}.
\end{enumerate}

\subsection{Fitting the SEDs}

For fitting the model, we use the least square method from \textsc{scipy} \texttt{curve\_fit} function \citep{Virtanen2020NatMe..17..261V}. First, we fit the source flux with a single-component blackbody curve. We tested various methods to determine a way to identify bad fits. The error-weighted residual (EWR) provided the simplest way to detect bad fits for a large range of parameters. 
\begin{equation}
    EWR = (F-F_{fit})/\epsilon_{F}
\end{equation}
where, $F$ is the observed flux, $F_{fit}$ is the fitted model flux and $\epsilon_{F}$ is the error in the flux.
We used a threshold of a minimum of 3 filters with $|EWR|>5$ to identify a \textit{bad} single fit (\texttt{threshold\_nfilter = 3} and \texttt{threshold\_ewr = 5}).
These bad single fits were then fitted with two-component blackbody using \texttt{curve\_fit}.

For simplicity, we assumed that $\sigma$ was the same in all filters except in Section~\ref{sec:case_errors} (where we have observed fractional errors in individual filters).
To estimate the errors in the stellar parameters, we repeated the same fitting process for 50 randomised spectra based on the specified $\sigma$ parameter. 
We defined an acceptable parameter space within $T$ of 1000--$10^6$ K and $L$ of 10$^{-10}$--10$^{12}$ L\sun. And then culled any failed fits and the fits near (within 1\%) or outside the boundary. For binary fits, extra culling was required due to frequent failed fits or extra deviations. Hence, a second culling was applied to remove any binary fits where the primary temperature had an error of $>$ 10\%.
We then defined the \texttt{convergence\_rate} as the ratio of the number of acceptable fits to total randomisations. The third culling was performed to enforce \texttt{convergence\_rate} of $\geq50$\%
\footnote{All these hyperparameters (number of randomisations, boundary conditions and culling thresholds) can be modified as per the user's needs.}. Only these \textit{good} fits are shown and discussed further.
The and 16th/84th percentile values of the converged fits were used as the recovered parameter and their errors, respectively.

The modelling and methodology mentioned here are combined into a \textsc{python} package \textsc{sed-analysis-tools}. The code\footnote{\url{https://github.com/jikrant3/sed-analysis-tools}}, documentation\footnote{\url{https://jikrant3.github.io/sed-analysis-tools}} and Jupyter notebooks required to recreate this work are available online \citep{jikrant3_2024_13789848}.

\vspace{-1em}
\section{Recoverability across the HRD} \label{sec:general_hrd}

\begin{figure}[ht]
    \centering
    \includegraphics[width=0.9\linewidth]{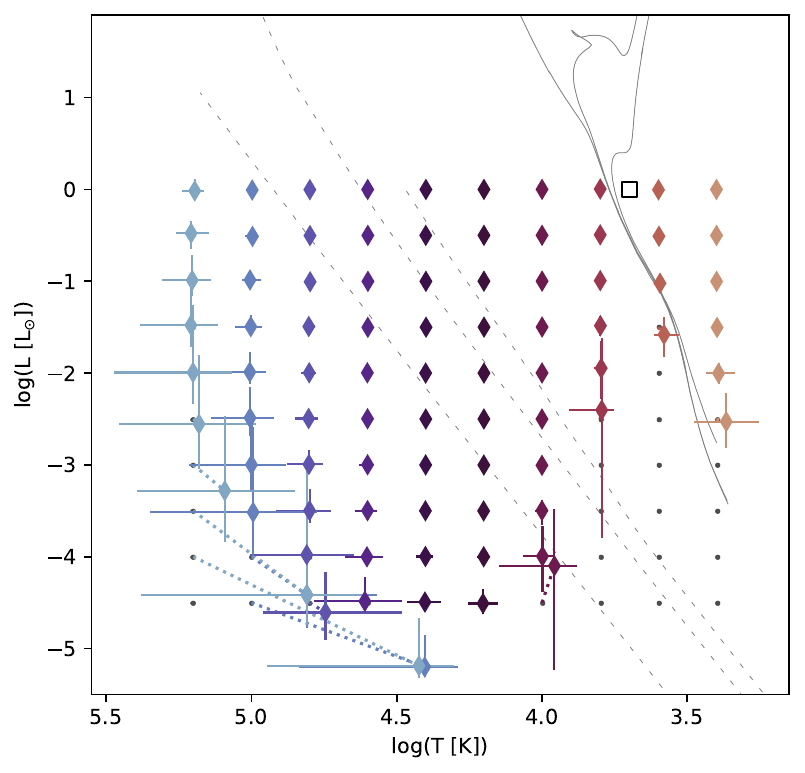}
    \caption{HRD of recovered secondary stars for a primary star of 5000 K and 1 L\sun\ modelled with the \texttt{optimistic\_filters} and 1\% flux error. The primary position is denoted by a black square. The isochrones (log($age$) $=8,9,10$; \citealt{Bressan2012MNRAS.427..127B}) and WD cooling curves (M$_{WD}=0.2,0.5,1.3$; \citealt{Tremblay2011ApJ...730..128T}) are shown as grey curves for reference. The priors of the secondaries are shown by grey dots while the recovered parameters of the secondaries are shown by diamonds coloured according to their prior temperature for visual clarity. The dotted lines connect the prior positions with the recovered median positions.}
    \label{fig:demo_hrd_grid}
\end{figure}

As pure blackbodies, the sources we consider have the same spectrum shape irrespective of the luminosity, and binaries with the same $L_1/L_2$ are equivalent to each other. Hence, we fixed the luminosity of the primary to be 1 L\sun. 
We created a list of primary sources with log($T$) in the range of 3.5 to 5.1 with steps of 0.2 dex. We added noise using log($\sigma$) within -4 to -1 with steps of 0.5 dex.

First, we tested the recoverability of the parameters of a single primary star. The fitting showed that the input parameters can be recovered satisfactorily for all positions within the HRD. For example, typical temperature errors were less than 5\% even for a spectrum with 50\% flux errors ($\sigma=0.5$). The single fits had significant errors primarily for stars whose peak wavelength falls outside the range of the SEDs (e.g. $T>50000$ K).
However, in reality, the stars are not simple blackbodies, thus the single fit might not be satisfactory in practice, depending on the assumptions about the spectral type, non-thermal emission, observational systematics, extinction and choice of spectral model.

Satisfied with the performance of single fits, we tested binary systems. We defined the binaries with the assumption that $L_1\geq L_2$. 
The tested primary stars have solar luminosity and log($T$) values of 3.5 to 5.1 with steps of 0.2 dex.
To test the recoverability of stellar parameters of various binary systems, we populated the secondary stars within a grid of log($T$) values from 3.4 to 5.2 in steps of 0.2 dex and log($L$) values of -4.5 to 0 in steps of 0.5 for each of the primary stars. 
We identified systems with unsatisfactory single fits and performed double component fitting.
 
Figure~\ref{fig:demo_hrd_grid} shows an example of the recovered parameters for a binary system with a primary of 5000 K and 1 L\sun. The faint and red secondaries had minimal effect on the overall SED and thus were not identifiable as binaries using the SED. The faint and very blue secondaries do change the shape of the single SED, but their recovered parameters had large errors (e.g. underestimation of temperature).
The faintest and hottest secondaries show convergence in two-component fits with accurate primary parameters. However, their secondary temperatures were significantly incorrect (although the errors calculated here cover the secondary priors within 2$\sigma$). For example, the secondary with log$T$ = 5.2 and log$L$ = $-$3.5 has similar recovered parameters as a secondary with log$T$ = 4.8 and log$L$ = $-$4.5. Thus, it is important to check for such degeneracy while analysing any results from two-component fits (see Section~\ref{sec:case_errors} for another example). Such degeneracy increased when the errors were large or the wavelength coverage was inadequate.

Figures~\ref{fig:hrd_set_optimistic}, and \ref{fig:hrd_set_realistic} show the HRD positions of the detected secondaries and the errors in their recovered temperatures for the \texttt{optimistic\_filters}, and \texttt{realistic\_filters} sets, respectively.

The \texttt{optimistic\_filters} set (Figure~\ref{fig:hrd_set_optimistic}) shows the detectable parameter space of the secondary stars. Hot primaries ($>$10000 K) dominated the flux in the UV range and thus suppressed the presence of hotter companions. In contrast, the cooler companions can be detected for a large temperature and luminosity range owing to their IR flux. 
The parameter recovery for the hottest primaries ($>75000$ K) with large errors (10\%) was poor because (i) the tail end of the blackbody spectrum has a poor constraint on the temperature and (ii) parameter estimation of one or both components fails due to higher noise.

The \texttt{realistic\_filters} case (Figure~\ref{fig:hrd_set_optimistic}) shows that with a realistic number of filters covering UV to mid-IR the recovery of secondaries was similar to the \texttt{optimistic\_filters}. The only major difference was a slight increase in the secondary errors. This shows that a well-covered broadband set of filters was enough for characterising binaries, and a large collection of narrowband filters might not improve the detectability significantly. We tested the same with 7 and 10 filters; however, the errors in stellar parameters increased (with noticeable offsets in median values) and the recovered region in the HRD shrunk significantly for fewer than 15 filters.

We also tested the performance of \texttt{gaia\_xp\_filters}, \texttt{uocs\_filters} and \texttt{optical\_IR\_filters} sets for the same binary grid. As expected, \texttt{uocs\_filters} results were very similar to \texttt{realistic\_filters} while \texttt{gaia\_xp\_filters} and \texttt{optical\_IR\_filters} recovered a smaller percentage of binaries due to their smaller wavelength coverage.

Overall, the detectability of the binary companion was strongly dependent on the primary temperature. The reliability of the secondary parameters was heavily dependent on noise. 

\vspace{-1em}
\section{Case Study for Detecting Unresolved Binaries with \textit{Gaia} XP Spectra}
\label{sec:case_unresolved}

\begin{figure}[ht]
    \centering
    \includegraphics[width=0.9\linewidth]{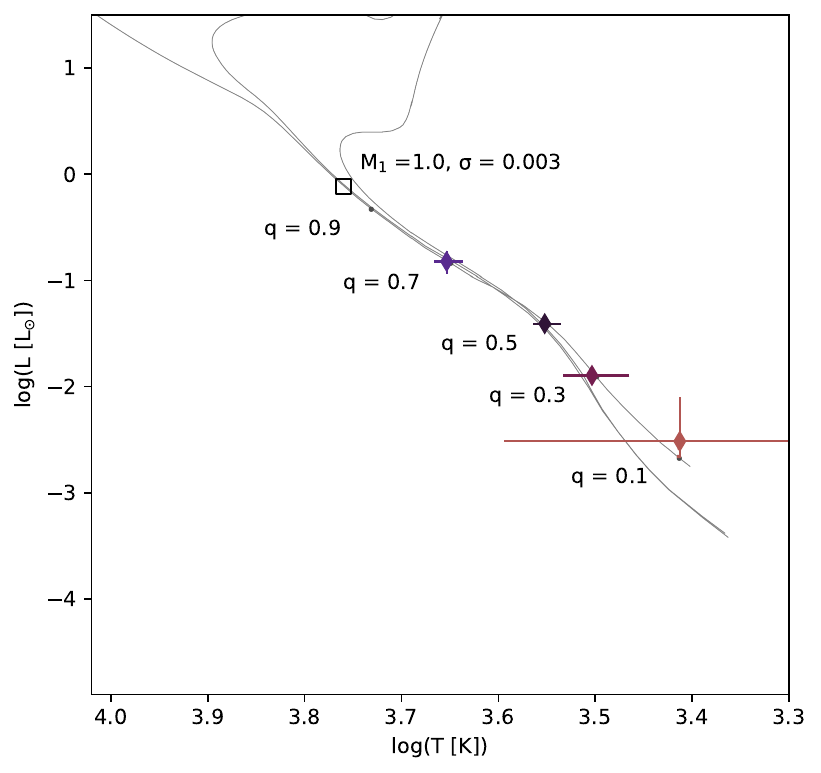}
    \caption{The HRD of recovered secondary stars for a primary of 1 M\sun\ modelled with the \texttt{gaia\_xp\_filters} and 0.3\% flux error. The markers are same as Figure~\ref{fig:demo_hrd_grid}}
    \label{fig:demo_hrd_unresolved}
\end{figure}

Figure~\ref{fig:demo_hrd_unresolved} shows an example of a 1 M\sun\ primary with companions of mass ratio ($q$) from 0.1 to 0.9. The results show that for such a primary (with 0.1\% flux error), one can only characterise secondaries with $q$ = 0.3, 0.5 and 0.7. 
The smallest secondary had large errors and the $q$ = 0.9 secondary has an indistinguishable spectrum at the given precision. 
The fitting is performed blindly in this analysis, hence we could not identify a binary with $q\approx1$ which is easily identifiable in simple colour-magnitude diagrams. An obvious improvement can be made by forcing the primaries and secondaries to lie on the MS. \citet{Thompson2021AJ....161..160T} used such a technique to successfully identify binaries with $q\in[0.1,1]$. They created discrete models for binaries with all mass ratios and fitted them with the observed data to obtain the best-fitting binary system. Analytically, one can modify Eq.~\ref{eq:flux_single} such that the radius is a function of temperature (based on the isochrones or theoretical relations) and perform the fits to enforce that the components follow the MS.

The results for a larger grid based on primaries, $\sigma$ and \texttt{gaia\_xp\_filters} are given in Figure~\ref{fig:hrd_set_unresolved_gaia}. 
In general, the figures show that it was easier to identify secondaries in low-mass primaries compared to high-mass primaries. 
The figure also shows that the majority of the unresolved binaries were detectable at the 1\% precision level of the typical \textit{Gaia} XP spectra, but their errors were too large.
Thus, only the lowest error and intermediate-mass ratio binaries would be characterisable using \textit{Gaia} XP spectra alone. Enforcing the MS nature of the binary components could be useful in increasing the recoverability as mentioned above. However, that is only possible in star clusters or with very accurate absolute magnitude estimates.

\vspace{-1em}
\section{Case Study for Detecting Hotter Companions Using UV-IR SED}
\label{sec:case_uocs}

\begin{figure}[ht]
    \centering
    \includegraphics[width=0.9\linewidth]{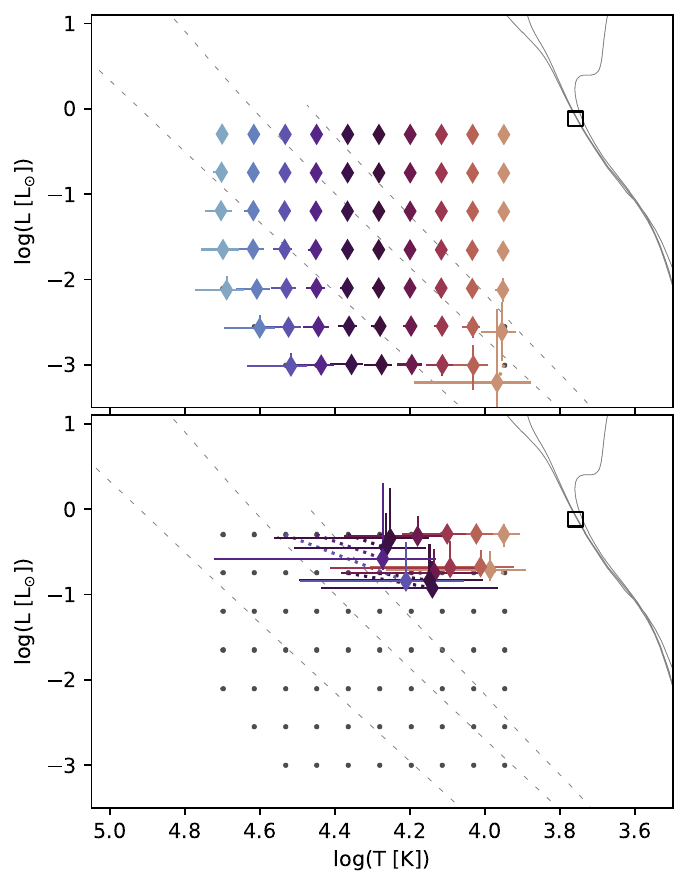}
    \caption{The HRD of recovered hotter companions for a primary of 1 M\sun\ modelled with the \texttt{uocs\_filters} (top) and \texttt{optical\_IR\_filters} (bottom), and 1\% flux error. The markers are same as Figure~\ref{fig:demo_hrd_grid}}
    \label{fig:demo_hrd_uocs}
\end{figure}

Recently, we have been conducting a UVIT Open cluster Study (UOCS) which focused on identifying and characterising hotter companions to optically bright stars in open clusters \citep{Sindhu2019ApJ...882...43S, Jadhav2023A&A...676A..47J, Panthi2024MNRAS.527.8325P, Pal2024ApJ...970L..39P}. To approximate the typical wavelength coverage in these studies, we used the \texttt{uocs\_filters} set. As a control experiment, we used the \texttt{optical\_IR\_filters} set which only covers the optical-IR range. The secondaries were populated according to the approximate region where hotter companions have been detected by UOCS studies \citep{Jadhav2024A&A...688A.152J}.

Figure~\ref{fig:demo_hrd_uocs} shows the recoverability of hotter companions with \texttt{uocs\_filters} and \texttt{optical\_IR\_filters} sets. The UV filters can detect the secondary and characterise it satisfactorily for most of the tested regime. The \texttt{optical\_IR\_filters} either completely missed the hotter companions or the parameterisation was too poor. 

Figure~\ref{fig:hrd_set_uocs} shows the recovery of hotter companions for primaries with masses from 0.8 to 3 M\sun. For 1\% flux errors, the hotter companions of low-mass primaries were easily recovered, however, the recovery regime of hotter companions was much smaller (in both $L$ and $T$) for higher-mass primaries ($>1.2$ M\sun). At 10\% flux errors, a very limited number of hotter companions can be recovered with accurate parameters.

Figure~\ref{fig:hrd_set_optIR} shows the hotter companion recovery for only optical-IR filters. And as expected, only the brightest hotter companions were recoverable. Due to the absence of UV filters, the flux from hotter companions was only detectable in case the flux errors were $\leq$1\% and for primaries with mass $\leq1.8$M\sun.

\vspace{-1em}
\section{Case study for Estimating Errors in Binary SED Fitting}
\label{sec:case_errors}

\begin{figure}[ht]
    \centering
    \includegraphics[width=0.9\linewidth]{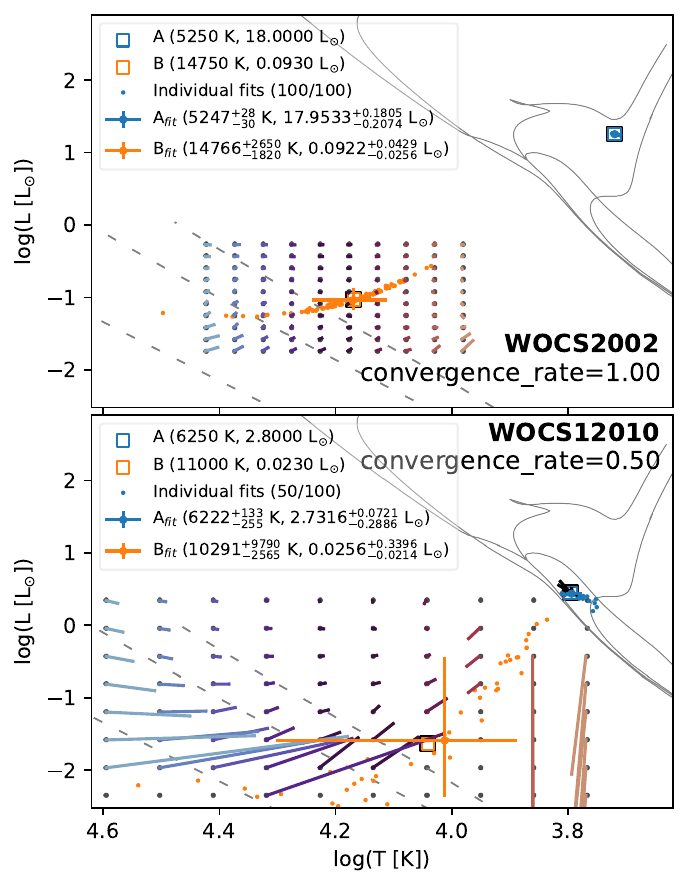}
    \caption{The HRD showing the error estimates for WOCS2002 (a known MS+WD system) and WOCS12010 (a MS+WD candidate) in open cluster M67 (\citealt{Jadhav2019ApJ...886...13J}; Pal et al. in prep). The prior positions are indicated by squares, while the recovered median positions are shown by circles with error bars. The individual random realisations are shown by orange (secondary) and blue (primary) dots. The grid of pseudo-secondaries is shown by grey dots. The similarly coloured lines connect each pseudo-secondary input position with its recovered median position.}
    \label{fig:wocs2002}
\end{figure}

Estimating the correct errors in binary SED fitting is quite complicated due to the choice of models, incorrectness in quoted flux errors, inadequate reddening corrections, and the fitting technique. One of the statistical way to calculate errors is bootstrapping; as implemented in \textsc{binary\_sed\_fitting} \citep{Jadhav2021JApA...42...89J} and \textsc{vosa} \citep{Bayo2008A&A...492..277B}. However, these estimates are calculated based on the successful fits and are burdened with all the assumptions about the stellar parameters and models. There is no analytical method to estimate the parameter errors. We propose using the results from a blackbody binary to get an estimate of the errors. This estimate can be considered as lower bound to the actual error because we assumed the simplest model (single/double component blackbody) and a well-defined noise.

Figure~\ref{fig:wocs2002} shows examples of error estimation using blackbody fitting. We used the results from the previous binary SED fitting to initialise a blackbody binary system with the same filters and the corresponding fractional errors. We then created 100 random realisations and calculated new median HRD position along with new estimations of errors.
\citet{Jadhav2019ApJ...886...13J} presented SED-based temperature values of WOCS 2002 as 5250$\pm$125 K and 14750$\pm$250 K based on a combination of bootstrapping and the model temperature grid. The current analysis shows that the recovered parameters would be $5247^{+28}_{-30}$ K and $14766^{+2650}_{-1820}$ K for the given precision. This shows that the mean values were correct in \citet{Jadhav2019ApJ...886...13J}, but the secondary parameter errors were severely underestimated.
Apart from the error estimates, binary SEDs also face the degeneracy problem mentioned in Section~\ref{sec:general_hrd}, i.e. whether any other secondary can have the same recovered parameters as WOCS2002-B. To understand the potential degeneracy, we created a grid of pseudo-secondaries near the HRD position of WOCS2002-B. The 0.05--0.95 percentile values of log$(T_2)$ and log$(L_2)$ of the random realisations were used as the boundaries of this grid.
The input and recovered positions of these pseudo-secondaries are shown connected by coloured lines in the figure. The figure shows that none of the other recovered pseudo-secondaries occupied the same position as the WOCS2002-B. Thus we concluded that at the given precision and filters, it was unlikely that any other secondary could mimic the WOCS2002-B properties and the errors were significantly larger than the typical model step size (e.g. 500 K for the Koester WD spectral models used in the previous analysis).

In contrast, the WOCS12010 example (Figure~\ref{fig:wocs2002} bottom panel) shows that its \texttt{convergence\_rate} is low and there is a chance that a fainter and hotter secondary can mimic the WOCS12010-B HRD position after SED fitting. Also, the WOCS12010-B HRD position is near the edge of the recoverable secondary region. Hence, caution is advised while dealing with such systems.

\vspace{-1em}
\section{Recommendations and limitations} \label{sec:reco}

As shown in previous sections, analysing blackbody binary SEDs can provide important information about the recoverability of the two components. We recommend performing a feasibility study using the expected noise levels, filter combinations and secondary parameter space for any binary SED analysis (similar to Section~\ref{sec:general_hrd}--\ref{sec:case_uocs}).

The user is recommended to keep the HRD acceptable boundary much larger than the expected stellar parameters. The number of random realisations should be $\geq50$ to get reliable error estimates. In the case of completely unknown systems, the user can forgo the constraint which forces the primary errors to be low and analyse the results accordingly. 
Depending on the data quality, the user can lower the \texttt{threshold\_nfilter} and \texttt{threshold\_ewr} to analyse the non-obvious binaries.
The \texttt{convergence\_rate} threshold was only used for plotting and summarising the generalized results. Overall, the best practice would be analysing the an individual fit based on the appropriate hyper-parameters.

The primary limitation of this approach is the approximation of all stars as blackbodies. One could model the stars more realistically using stellar spectral models which would provide better results. However, most available spectral libraries are tailored to particular types of stars and their parameter space is limited to the assumed stellar types. Also, in the case of unknown components, it is difficult to choose the spectral library a priori. To avoid the limits of parameter space, stellar types and inherent uncertainties in the spectral libraries, we have used blackbody modelling as the best approach. If a binary system has known stellar types and suitable spectral models, one can use the same methodology to get better error estimates and recoverability. However, such a targeted analysis is beyond the scope of the current work.

\vspace{-1em}
\section{Summary and Conclusions} \label{sec:conclusions}

We analysed the recoverability of binary parameters using multiwavelength SEDs spanning UV to mid-IR region. We assumed that the stars were pure blackbodies and tested a variety of primaries, secondaries, filter combinations and photometric uncertainties to identify the recoverable regime in the HRD. We also measured the accuracy of the recovered parameters. Based on the analysis, we reach the following conclusions.

\begin{itemize}
    \item The noisy ($\geq$10\% flux error) sources hotter than 50000 K were difficult to fit using SEDs as their peak wavelength is beyond the UV range. The same was true for binaries consisting of such noisy and hot stars.
    \item The parameter recovery using 15 and 50 filters in the UV-IR range has a negligible difference. Thus, a collection of medium/wide-bands covering UV-IR is enough for binary parameterisation and including narrow-band filters is not necessary to improve the recovery of $T$ and $L$ (in the context of straightforward SED fitting without considering specific spectral features). We recommend using $\gtrapprox$15 filters to fit two-component models.
    \item The \textit{Gaia} XP spectra provide poor constraints on individual MS-MS binary systems except for the most precise and unequal mass ($q\approx0.3-0.7$) binaries. Enforcing the components to follow the MS could help increase the recoverability.
    \item UV flux was important to identify compact hotter companions to MS-type stars. However, the recovery and accuracy depend heavily on the primary temperature and photometric precision.
    \item Estimations of stellar parameters based on statistical methods (such as bootstrapping) and model grid steps are sometimes severely underestimated. We provide a recipe to get lower bounds on the fitting errors based on blackbody models. Unfortunately, an analytical prescription for predicting the $T$ and $L$ uncertainties for a given set of filters, errors and binary parameters was not possible. 
    \item In addition to errors, the degeneracy of various secondaries having similar fitting parameters needs to be accounted for before trusting the results of an SED fit.
\end{itemize}

Overall, the figures presented here and the related \textsc{python} package \href{https://github.com/jikrant3/sed-analysis-tools}{\textsc{sed-analysis-tools}} provide basic tools to gauge if a particular binary system is recoverable for a given filter system and estimate their uncertainty limits based on the photometric errors.

\appendix

\section{Supplementary figures}

\begin{figure*}
    \centering
    \includegraphics[width=0.98\linewidth]{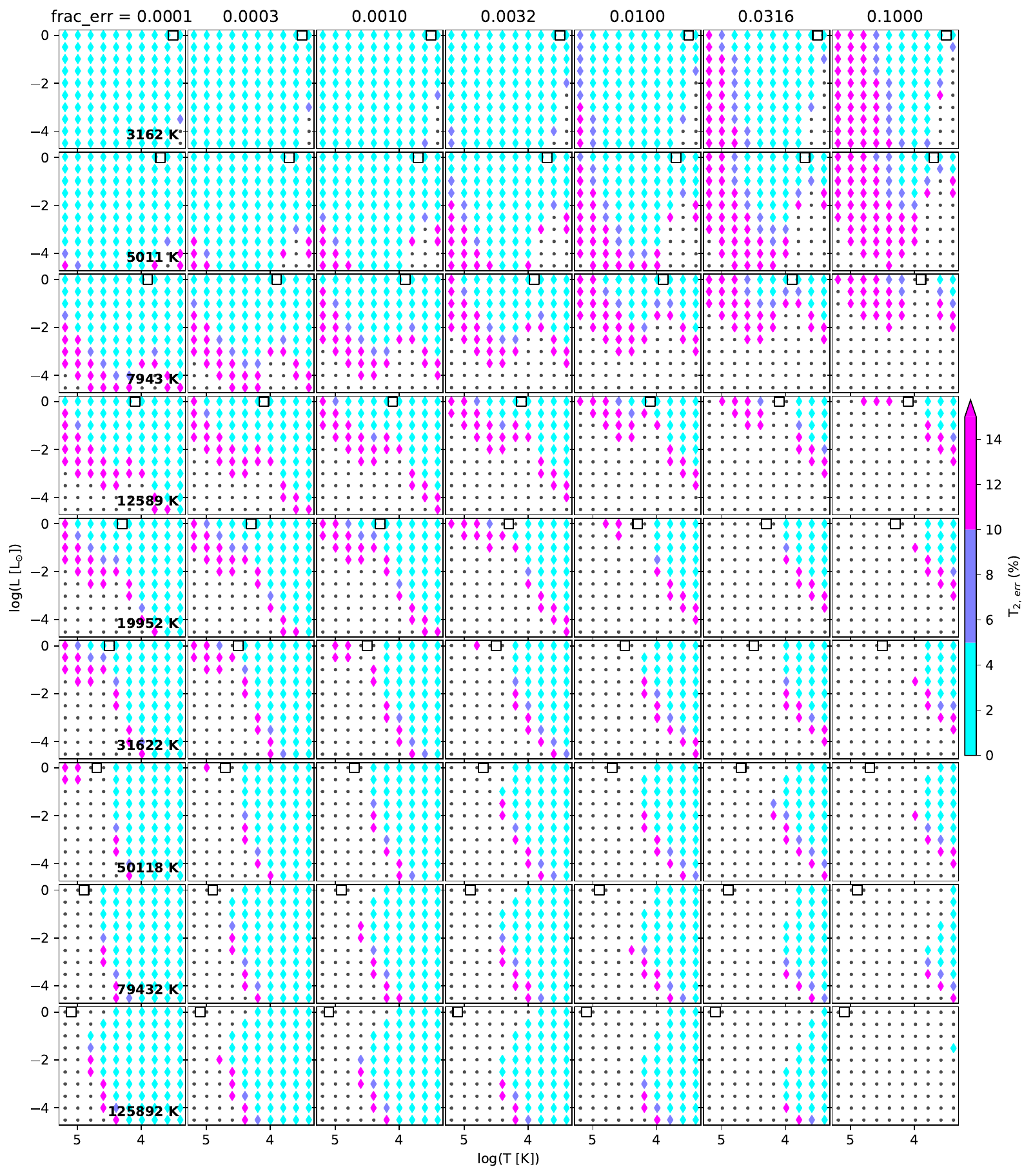}
    \caption{HRD positions of secondary stars for various primary stars using \texttt{optimistic\_filters}. The columns show results for increasing errors ($\sigma\in[0.0001,0.1]$), while the rows show errors for increasing primary temperature ($T\in[3162,125892 K]$). The primary position is denoted by a black square. All the secondaries are shown by grey dots. The recovered secondaries (diamonds) are coloured according to the percentage error in secondary temperature ($T_{2,\ err}$).}
    \label{fig:hrd_set_optimistic}
\end{figure*}

\begin{figure*}
    \centering
    \includegraphics[width=0.98\linewidth]{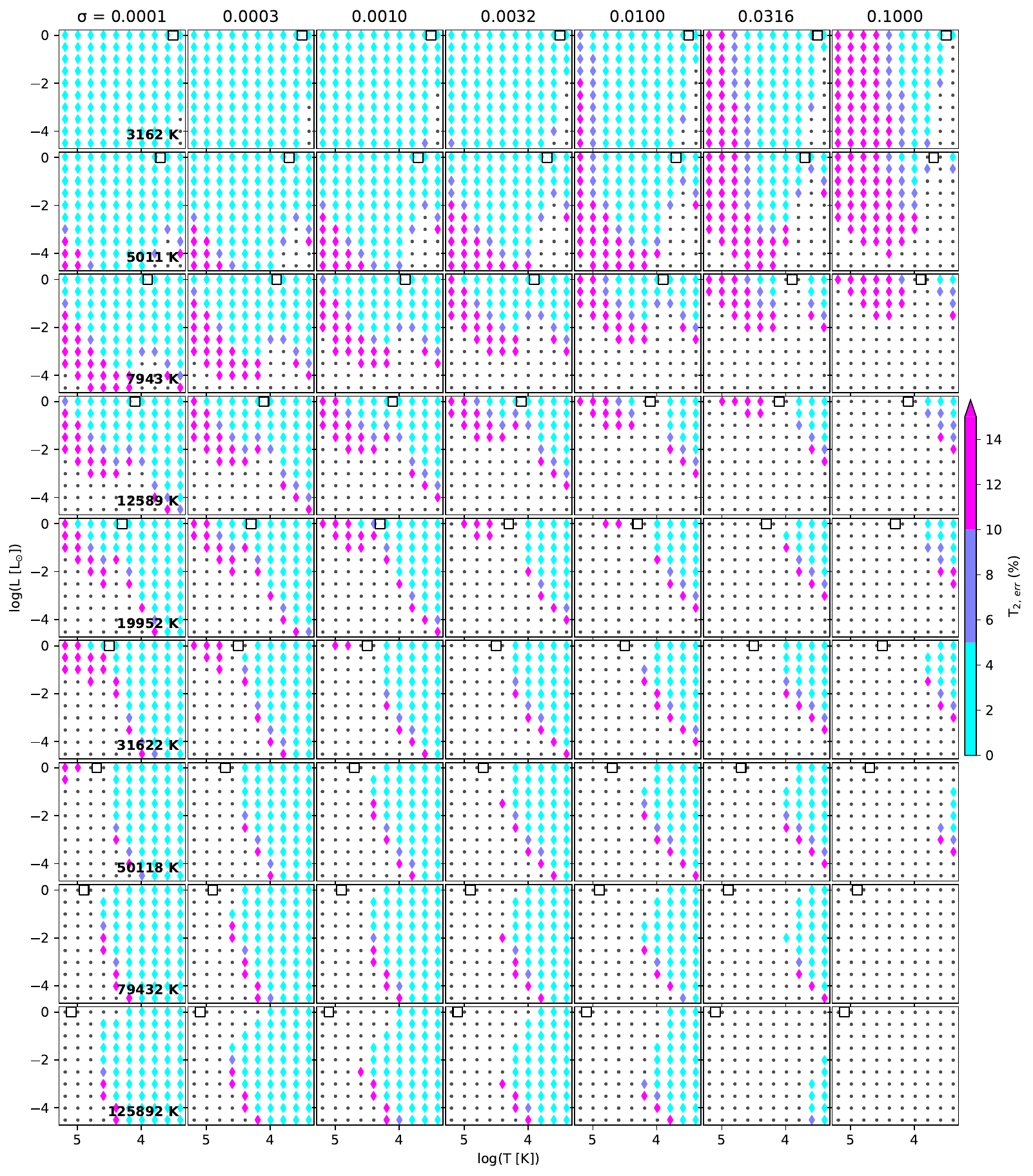}
    \caption{HRD positions of secondary stars for various primary stars using \texttt{realistic\_filters}. All other properties are similar to Figure~\ref{fig:hrd_set_optimistic}.}
    \label{fig:hrd_set_realistic}
\end{figure*}

\begin{figure*}
    \centering
    \includegraphics[width=0.98\linewidth]{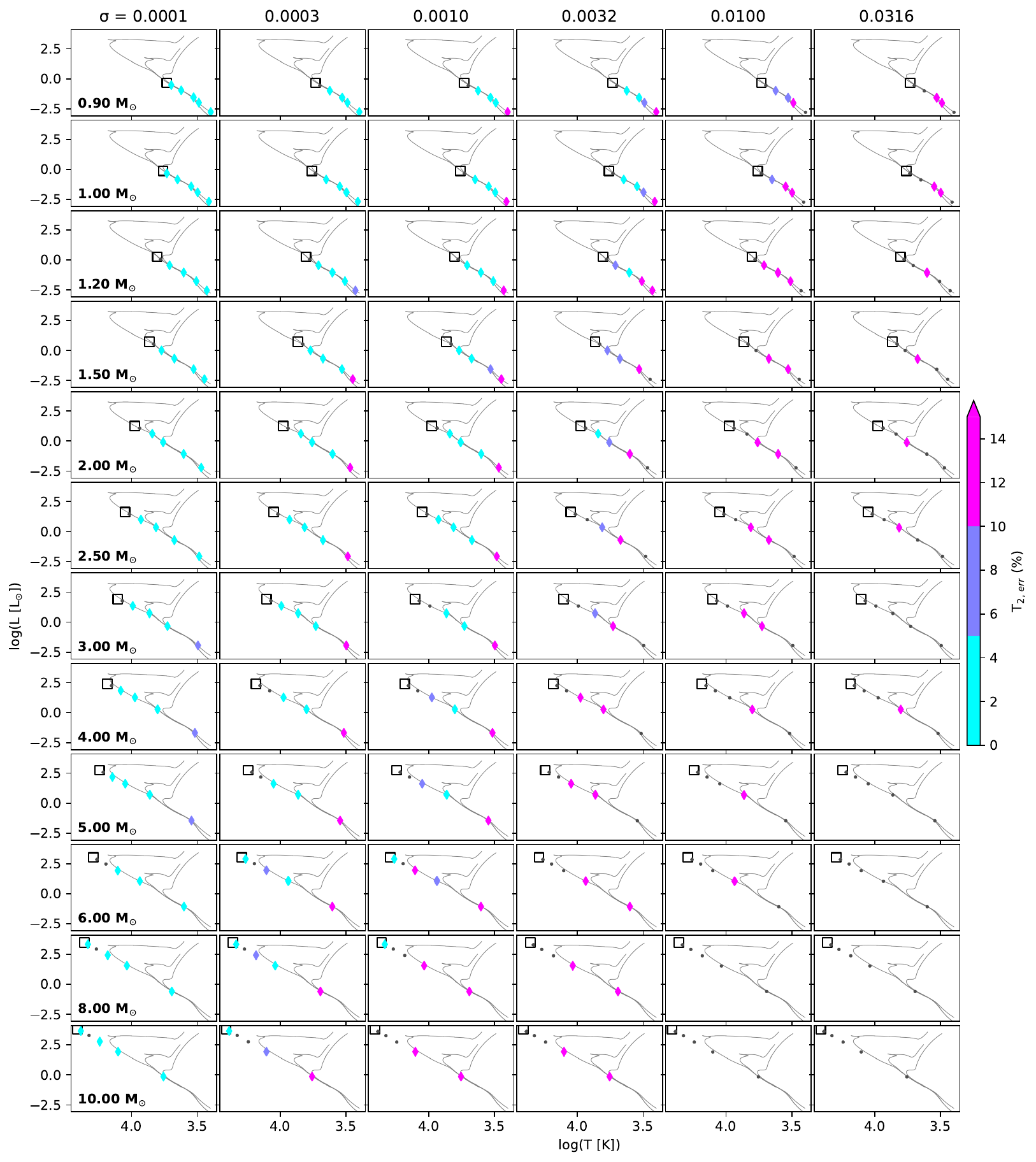}
    \caption{HRD positions of secondary stars (for $q\in[0.1,0.9]$) for various primary stars (M$_1\in[0.9,10\ M_{\odot}]$) using \texttt{gaia\_xp\_filters}. All other properties are similar to Figure~\ref{fig:hrd_set_optimistic}.}
    \label{fig:hrd_set_unresolved_gaia}
\end{figure*}

\begin{figure*}
    \centering
    \includegraphics[width=0.98\linewidth]{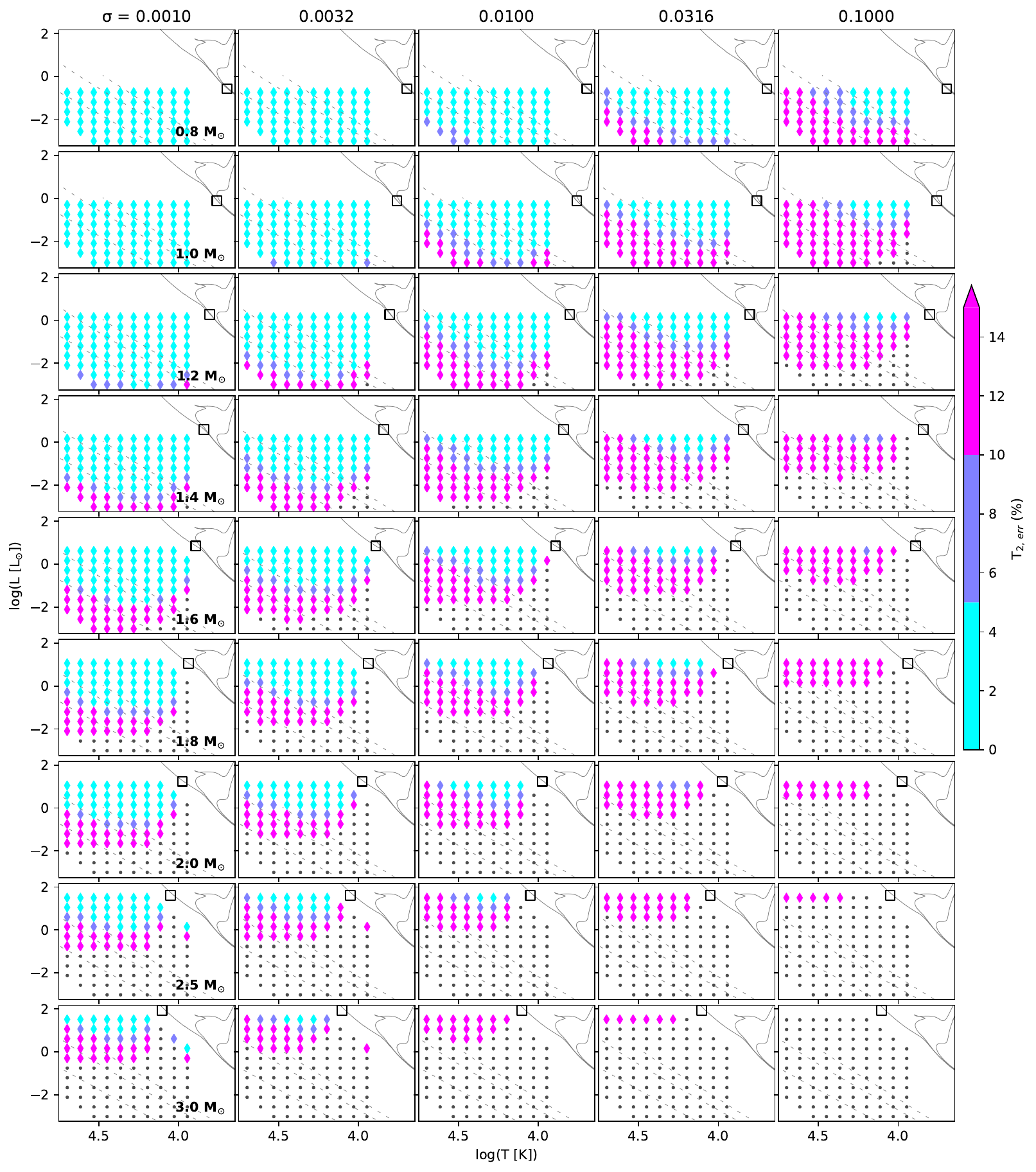}
    \caption{HRD positions of hotter companions for various primary stars (M$_1\in[0.8,3\ M_{\odot}]$) using \texttt{uocs\_filters}. All other properties are similar to Figure~\ref{fig:hrd_set_optimistic}.}
    \label{fig:hrd_set_uocs}
\end{figure*}

\begin{figure*}
    \centering
    \includegraphics[width=0.98\linewidth]{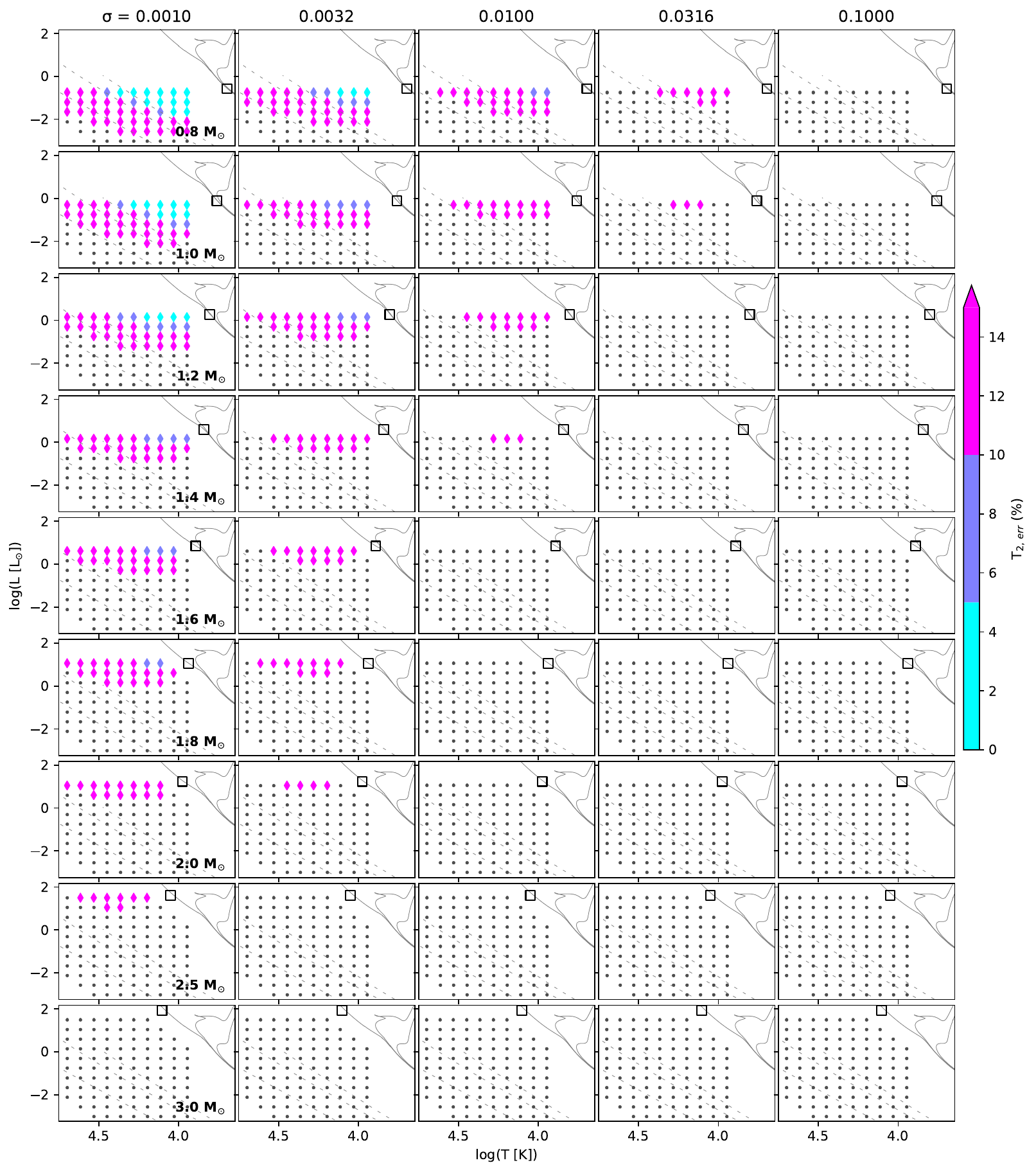}
    \caption{HRD positions of hotter companions for various primary stars (M$_1\in[0.8,3\ M_{\odot}]$) using \texttt{optical\_IR\_filters}. All other properties are similar to Figure~\ref{fig:hrd_set_optimistic}.}
    \label{fig:hrd_set_optIR}
\end{figure*}

\section*{Acknowledgements}
We thank the anonymous referees for the constructive comments. 
We thank A. Bayo for interesting discussion and improvement suggestions.
VJ thanks the Alexander von Humboldt Foundation for their support. The data analysis was done using \textsc{python} modules \textsc{astropy} \citep{Astropy2013A&A...558A..33A}, \textsc{numpy} \citep{harris2020array} and \textsc{scipy} \citep{Oliphant2007CSE.....9c..10O}, \textsc{pandas} \citep{pandas2022zndo...3509134T}, \textsc{matplotlib} \citep{Hunter2007CSE.....9...90H}, \textsc{ipython} \citep{Perez2007CSE.....9c..21P} and \textsc{sed-analysis-tools} \citep{jikrant3_2024_13789848}.
\vspace{-1em}


\bibliography{references}

\begin{thebibliography}{}
\expandafter\ifx\csname natexlab\endcsname\relax\def\natexlab#1{#1}\fi

\bibitem[{{Allard} {$et~al$.}(2003){Allard}, {Guillot}, {Ludwig}, {Hauschildt}, {Schweitzer}, {Alexander}, \& {Ferguson}}]{Allard2003IAUS..211..325A}
{Allard}, F., {Guillot}, T., {Ludwig}, H.-G., {$et~al$.} 2003, in IAU Symposium, Vol. 211, Brown Dwarfs, ed. E.~{Mart{\'\i}n}, 325

\bibitem[{{Allard} {$et~al$.}(2001){Allard}, {Hauschildt}, {Alexander}, {Tamanai}, \& {Schweitzer}}]{Allard2001ApJ...556..357A}
{Allard}, F., {Hauschildt}, P.~H., {Alexander}, D.~R., {Tamanai}, A., \& {Schweitzer}, A. 2001, \apj, 556, 357

\bibitem[{{Allard} {$et~al$.}(2011){Allard}, {Homeier}, \& {Freytag}}]{Allard2011ASPC..448...91A}
{Allard}, F., {Homeier}, D., \& {Freytag}, B. 2011, in Astronomical Society of the Pacific Conference Series, Vol. 448, 16th Cambridge Workshop on Cool Stars, Stellar Systems, and the Sun, ed. C.~{Johns-Krull}, M.~K. {Browning}, \& A.~A. {West}, 91

\bibitem[{{Andrae}(2010)}]{Andrae2010arXiv1009.2755A}
{Andrae}, R. 2010, arXiv e-prints, arXiv:1009.2755

\bibitem[{{Andrae} {$et~al$.}(2010){Andrae}, {Schulze-Hartung}, \& {Melchior}}]{Andrae2010arXiv1012.3754A}
{Andrae}, R., {Schulze-Hartung}, T., \& {Melchior}, P. 2010, arXiv e-prints, arXiv:1012.3754

\bibitem[{{Astropy Collaboration} {$et~al$.}(2013){Astropy Collaboration}, {Robitaille}, {Tollerud}, {Greenfield}, {Droettboom}, {Bray}, {Aldcroft}, {Davis}, {Ginsburg}, {Price-Whelan}, {Kerzendorf}, {Conley}, {Crighton}, {Barbary}, {Muna}, {Ferguson}, {Grollier}, {Parikh}, {Nair}, {Unther}, {Deil}, {Woillez}, {Conseil}, {Kramer}, {Turner}, {Singer}, {Fox}, {Weaver}, {Zabalza}, {Edwards}, {Azalee Bostroem}, {Burke}, {Casey}, {Crawford}, {Dencheva}, {Ely}, {Jenness}, {Labrie}, {Lim}, {Pierfederici}, {Pontzen}, {Ptak}, {Refsdal}, {Servillat}, \& {Streicher}}]{Astropy2013A&A...558A..33A}
{Astropy Collaboration}, {Robitaille}, T.~P., {Tollerud}, E.~J., {$et~al$.} 2013, \aap, 558, A33

\bibitem[{{Bayo} {$et~al$.}(2008){Bayo}, {Rodrigo}, {Barrado Y Navascu{\'e}s}, {Solano}, {Guti{\'e}rrez}, {Morales-Calder{\'o}n}, \& {Allard}}]{Bayo2008A&A...492..277B}
{Bayo}, A., {Rodrigo}, C., {Barrado Y Navascu{\'e}s}, D., {$et~al$.} 2008, \aap, 492, 277

\bibitem[{{Boquien} {$et~al$.}(2019){Boquien}, {Burgarella}, {Roehlly}, {Buat}, {Ciesla}, {Corre}, {Inoue}, \& {Salas}}]{Boquien2019A&A...622A.103B}
{Boquien}, M., {Burgarella}, D., {Roehlly}, Y., {$et~al$.} 2019, \aap, 622, A103

\bibitem[{{Bressan} {$et~al$.}(2012){Bressan}, {Marigo}, {Girardi}, {Salasnich}, {Dal Cero}, {Rubele}, \& {Nanni}}]{Bressan2012MNRAS.427..127B}
{Bressan}, A., {Marigo}, P., {Girardi}, L., {$et~al$.} 2012, \mnras, 427, 127

\bibitem[{{Castelli} \& {Kurucz}(2003)}]{Castelli2003IAUS..210P.A20C}
{Castelli}, F., \& {Kurucz}, R.~L. 2003, in IAU Symposium, Vol. 210, Modelling of Stellar Atmospheres, ed. N.~{Piskunov}, W.~W. {Weiss}, \& D.~F. {Gray}, A20

\bibitem[{{Coelho}(2014)}]{Coelho2014MNRAS.440.1027C}
{Coelho}, P.~R.~T. 2014, \mnras, 440, 1027

\bibitem[{{Conroy} \& {Gunn}(2010)}]{Conroy2010ApJ...712..833C}
{Conroy}, C., \& {Gunn}, J.~E. 2010, \apj, 712, 833

\bibitem[{{Doi} {$et~al$.}(2010){Doi}, {Tanaka}, {Fukugita}, {Gunn}, {Yasuda}, {Ivezi{\'c}}, {Brinkmann}, {de Haars}, {Kleinman}, {Krzesinski}, \& {French Leger}}]{Doi2010AJ....139.1628D}
{Doi}, M., {Tanaka}, M., {Fukugita}, M., {$et~al$.} 2010, \aj, 139, 1628

\bibitem[{{Gaia Collaboration} {$et~al$.}(2016){Gaia Collaboration}, {Prusti}, {de Bruijne}, {Brown}, {Vallenari}, {Babusiaux}, {Bailer-Jones}, {Bastian}, {Biermann}, {Evans}, {Eyer}, {Jansen}, {Jordi}, {Klioner}, {Lammers}, {Lindegren}, {Luri}, {Mignard}, {Milligan}, {Panem}, {Poinsignon}, {Pourbaix}, {Randich}, {Sarri}, {Sartoretti}, {Siddiqui}, {Soubiran}, {Valette}, {van Leeuwen}, {Walton}, {Aerts}, {Arenou}, {Cropper}, {Drimmel}, {H{\o}g}, {Katz}, {Lattanzi}, {O'Mullane}, {Grebel}, {Holland}, {Huc}, {Passot}, {Bramante}, {Cacciari}, {Casta{\~n}eda}, {Chaoul}, {Cheek}, {De Angeli}, {Fabricius}, {Guerra}, {Hern{\'a}ndez}, {Jean-Antoine-Piccolo}, {Masana}, {Messineo}, {Mowlavi}, {Nienartowicz}, {Ord{\'o}{\~n}ez-Blanco}, {Panuzzo}, {Portell}, {Richards}, {Riello}, {Seabroke}, {Tanga}, {Th{\'e}venin}, {Torra}, {Els}, {Gracia-Abril}, {Comoretto}, {Garcia-Reinaldos}, {Lock}, {Mercier}, {Altmann}, {Andrae}, {Astraatmadja}, {Bellas-Velidis}, {Benson}, {Berthier}, {Blomme}, {Busso}, {Carry}, {Cellino},
  {Clementini}, {Cowell}, {Creevey}, {Cuypers}, {Davidson}, {De Ridder}, {de Torres}, {Delchambre}, {Dell'Oro}, {Ducourant}, {Fr{\'e}mat}, {Garc{\'\i}a-Torres}, {Gosset}, {Halbwachs}, {Hambly}, {Harrison}, {Hauser}, {Hestroffer}, {Hodgkin}, {Huckle}, {Hutton}, {Jasniewicz}, {Jordan}, {Kontizas}, {Korn}, {Lanzafame}, {Manteiga}, {Moitinho}, {Muinonen}, {Osinde}, {Pancino}, {Pauwels}, {Petit}, {Recio-Blanco}, {Robin}, {Sarro}, {Siopis}, {Smith}, {Smith}, {Sozzetti}, {Thuillot}, {van Reeven}, {Viala}, {Abbas}, {Abreu Aramburu}, {Accart}, {Aguado}, {Allan}, {Allasia}, {Altavilla}, {{\'A}lvarez}, {Alves}, {Anderson}, {Andrei}, {Anglada Varela}, {Antiche}, {Antoja}, {Ant{\'o}n}, {Arcay}, {Atzei}, {Ayache}, {Bach}, {Baker}, {Balaguer-N{\'u}{\~n}ez}, {Barache}, {Barata}, {Barbier}, {Barblan}, {Baroni}, {Barrado y Navascu{\'e}s}, {Barros}, {Barstow}, {Becciani}, {Bellazzini}, {Bellei}, {Bello Garc{\'\i}a}, {Belokurov}, {Bendjoya}, {Berihuete}, {Bianchi}, {Bienaym{\'e}}, {Billebaud}, {Blagorodnova}, {Blanco-Cuaresma},
  {Boch}, {Bombrun}, {Borrachero}, {Bouquillon}, {Bourda}, {Bouy}, {Bragaglia}, {Breddels}, {Brouillet}, {Br{\"u}semeister}, {Bucciarelli}, {Budnik}, {Burgess}, {Burgon}, {Burlacu}, {Busonero}, {Buzzi}, {Caffau}, {Cambras}, {Campbell}, {Cancelliere}, {Cantat-Gaudin}, {Carlucci}, {Carrasco}, {Castellani}, {Charlot}, {Charnas}, {Charvet}, {Chassat}, {Chiavassa}, {Clotet}, {Cocozza}, {Collins}, {Collins}, {Costigan}, {Crifo}, {Cross}, {Crosta}, {Crowley}, {Dafonte}, {Damerdji}, {Dapergolas}, {David}, {David}, {De Cat}, {de Felice}, {de Laverny}, {De Luise}, {De March}, {de Martino}, {de Souza}, {Debosscher}, {del Pozo}, {Delbo}, {Delgado}, {Delgado}, {di Marco}, {Di Matteo}, {Diakite}, {Distefano}, {Dolding}, {Dos Anjos}, {Drazinos}, {Dur{\'a}n}, {Dzigan}, {Ecale}, {Edvardsson}, {Enke}, {Erdmann}, {Escolar}, {Espina}, {Evans}, {Eynard Bontemps}, {Fabre}, {Fabrizio}, {Faigler}, {Falc{\~a}o}, {Farr{\`a}s Casas}, {Faye}, {Federici}, {Fedorets}, {Fern{\'a}ndez-Hern{\'a}ndez}, {Fernique}, {Fienga}, {Figueras},
  {Filippi}, {Findeisen}, {Fonti}, {Fouesneau}, {Fraile}, {Fraser}, {Fuchs}, {Furnell}, {Gai}, {Galleti}, {Galluccio}, {Garabato}, {Garc{\'\i}a-Sedano}, {Gar{\'e}}, {Garofalo}, {Garralda}, {Gavras}, {Gerssen}, {Geyer}, {Gilmore}, {Girona}, {Giuffrida}, {Gomes}, {Gonz{\'a}lez-Marcos}, {Gonz{\'a}lez-N{\'u}{\~n}ez}, {Gonz{\'a}lez-Vidal}, {Granvik}, {Guerrier}, {Guillout}, {Guiraud}, {G{\'u}rpide}, {Guti{\'e}rrez-S{\'a}nchez}, {Guy}, {Haigron}, {Hatzidimitriou}, {Haywood}, {Heiter}, {Helmi}, {Hobbs}, {Hofmann}, {Holl}, {Holland}, {Hunt}, {Hypki}, {Icardi}, {Irwin}, {Jevardat de Fombelle}, {Jofr{\'e}}, {Jonker}, {Jorissen}, {Julbe}, {Karampelas}, {Kochoska}, {Kohley}, {Kolenberg}, {Kontizas}, {Koposov}, {Kordopatis}, {Koubsky}, {Kowalczyk}, {Krone-Martins}, {Kudryashova}, {Kull}, {Bachchan}, {Lacoste-Seris}, {Lanza}, {Lavigne}, {Le Poncin-Lafitte}, {Lebreton}, {Lebzelter}, {Leccia}, {Leclerc}, {Lecoeur-Taibi}, {Lemaitre}, {Lenhardt}, {Leroux}, {Liao}, {Licata}, {Lindstr{\o}m}, {Lister}, {Livanou}, {Lobel},
  {L{\"o}ffler}, {L{\'o}pez}, {Lopez-Lozano}, {Lorenz}, {Loureiro}, {MacDonald}, {Magalh{\~a}es Fernandes}, {Managau}, {Mann}, {Mantelet}, {Marchal}, {Marchant}, {Marconi}, {Marie}, {Marinoni}, {Marrese}, {Marschalk{\'o}}, {Marshall}, {Mart{\'\i}n-Fleitas}, {Martino}, {Mary}, {Matijevi{\v{c}}}, {Mazeh}, {McMillan}, {Messina}, {Mestre}, {Michalik}, {Millar}, {Miranda}, {Molina}, {Molinaro}, {Molinaro}, {Moln{\'a}r}, {Moniez}, {Montegriffo}, {Monteiro}, {Mor}, {Mora}, {Morbidelli}, {Morel}, {Morgenthaler}, {Morley}, {Morris}, {Mulone}, {Muraveva}, {Musella}, {Narbonne}, {Nelemans}, {Nicastro}, {Noval}, {Ord{\'e}novic}, {Ordieres-Mer{\'e}}, {Osborne}, {Pagani}, {Pagano}, {Pailler}, {Palacin}, {Palaversa}, {Parsons}, {Paulsen}, {Pecoraro}, {Pedrosa}, {Pentik{\"a}inen}, {Pereira}, {Pichon}, {Piersimoni}, {Pineau}, {Plachy}, {Plum}, {Poujoulet}, {Pr{\v{s}}a}, {Pulone}, {Ragaini}, {Rago}, {Rambaux}, {Ramos-Lerate}, {Ranalli}, {Rauw}, {Read}, {Regibo}, {Renk}, {Reyl{\'e}}, {Ribeiro}, {Rimoldini}, {Ripepi}, {Riva},
  {Rixon}, {Roelens}, {Romero-G{\'o}mez}, {Rowell}, {Royer}, {Rudolph}, {Ruiz-Dern}, {Sadowski}, {Sagrist{\`a} Sell{\'e}s}, {Sahlmann}, {Salgado}, {Salguero}, {Sarasso}, {Savietto}, {Schnorhk}, {Schultheis}, {Sciacca}, {Segol}, {Segovia}, {Segransan}, {Serpell}, {Shih}, {Smareglia}, {Smart}, {Smith}, {Solano}, {Solitro}, {Sordo}, {Soria Nieto}, {Souchay}, {Spagna}, {Spoto}, {Stampa}, {Steele}, {Steidelm{\"u}ller}, {Stephenson}, {Stoev}, {Suess}, {S{\"u}veges}, {Surdej}, {Szabados}, {Szegedi-Elek}, {Tapiador}, {Taris}, {Tauran}, {Taylor}, {Teixeira}, {Terrett}, {Tingley}, {Trager}, {Turon}, {Ulla}, {Utrilla}, {Valentini}, {van Elteren}, {Van Hemelryck}, {van Leeuwen}, {Varadi}, {Vecchiato}, {Veljanoski}, {Via}, {Vicente}, {Vogt}, {Voss}, {Votruba}, {Voutsinas}, {Walmsley}, {Weiler}, {Weingrill}, {Werner}, {Wevers}, {Whitehead}, {Wyrzykowski}, {Yoldas}, {{\v{Z}}erjal}, {Zucker}, {Zurbach}, {Zwitter}, {Alecu}, {Allen}, {Allende Prieto}, {Amorim}, {Anglada-Escud{\'e}}, {Arsenijevic}, {Azaz}, {Balm}, {Beck},
  {Bernstein}, {Bigot}, {Bijaoui}, {Blasco}, {Bonfigli}, {Bono}, {Boudreault}, {Bressan}, {Brown}, {Brunet}, {Bunclark}, {Buonanno}, {Butkevich}, {Carret}, {Carrion}, {Chemin}, {Ch{\'e}reau}, {Corcione}, {Darmigny}, {de Boer}, {de Teodoro}, {de Zeeuw}, {Delle Luche}, {Domingues}, {Dubath}, {Fodor}, {Fr{\'e}zouls}, {Fries}, {Fustes}, {Fyfe}, {Gallardo}, {Gallegos}, {Gardiol}, {Gebran}, {Gomboc}, {G{\'o}mez}, {Grux}, {Gueguen}, {Heyrovsky}, {Hoar}, {Iannicola}, {Isasi Parache}, {Janotto}, {Joliet}, {Jonckheere}, {Keil}, {Kim}, {Klagyivik}, {Klar}, {Knude}, {Kochukhov}, {Kolka}, {Kos}, {Kutka}, {Lainey}, {LeBouquin}, {Liu}, {Loreggia}, {Makarov}, {Marseille}, {Martayan}, {Martinez-Rubi}, {Massart}, {Meynadier}, {Mignot}, {Munari}, {Nguyen}, {Nordlander}, {Ocvirk}, {O'Flaherty}, {Olias Sanz}, {Ortiz}, {Osorio}, {Oszkiewicz}, {Ouzounis}, {Palmer}, {Park}, {Pasquato}, {Peltzer}, {Peralta}, {P{\'e}turaud}, {Pieniluoma}, {Pigozzi}, {Poels}, {Prat}, {Prod'homme}, {Raison}, {Rebordao}, {Risquez}, {Rocca-Volmerange},
  {Rosen}, {Ruiz-Fuertes}, {Russo}, {Sembay}, {Serraller Vizcaino}, {Short}, {Siebert}, {Silva}, {Sinachopoulos}, {Slezak}, {Soffel}, {Sosnowska}, {Strai{\v{z}}ys}, {ter Linden}, {Terrell}, {Theil}, {Tiede}, {Troisi}, {Tsalmantza}, {Tur}, {Vaccari}, {Vachier}, {Valles}, {Van Hamme}, {Veltz}, {Virtanen}, {Wallut}, {Wichmann}, {Wilkinson}, {Ziaeepour}, \& {Zschocke}}]{Gaia2016A&A...595A...1G}
{Gaia Collaboration}, {Prusti}, T., {de Bruijne}, J.~H.~J., {$et~al$.} 2016, \aap, 595, A1

\bibitem[{{Gosnell} {$et~al$.}(2015){Gosnell}, {Mathieu}, {Geller}, {Sills}, {Leigh}, \& {Knigge}}]{Gosnell2015ApJ...814..163G}
{Gosnell}, N.~M., {Mathieu}, R.~D., {Geller}, A.~M., {$et~al$.} 2015, \apj, 814, 163

\bibitem[{Harris {$et~al$.}(2020)Harris, Millman, van~der Walt, Gommers, Virtanen, Cournapeau, Wieser, Taylor, Berg, Smith, Kern, Picus, Hoyer, van Kerkwijk, Brett, Haldane, del R{\'{i}}o, Wiebe, Peterson, G{\'{e}}rard-Marchant, Sheppard, Reddy, Weckesser, Abbasi, Gohlke, \& Oliphant}]{harris2020array}
Harris, C.~R., Millman, K.~J., van~der Walt, S.~J., {$et~al$.} 2020, Nature, 585, 357

\bibitem[{{Hunter}(2007)}]{Hunter2007CSE.....9...90H}
{Hunter}, J.~D. 2007, Computing in Science and Engineering, 9, 90

\bibitem[{{Jadhav} {$et~al$.}(2024){Jadhav}, {Subramaniam}, \& {Sagar}}]{Jadhav2024A&A...688A.152J}
{Jadhav}, V., {Subramaniam}, A., \& {Sagar}, R. 2024, \aap, 688, A152

\bibitem[{{Jadhav}(2024)}]{jikrant3_2024_13789848}
{Jadhav}, V.~V. 2024, jikrant3/sed-analysis-tools, doi:10.5281/zenodo.13789847

\bibitem[{{Jadhav} {$et~al$.}(2021){Jadhav}, {Pandey}, {Subramaniam}, \& {Sagar}}]{Jadhav2021JApA...42...89J}
{Jadhav}, V.~V., {Pandey}, S., {Subramaniam}, A., \& {Sagar}, R. 2021, Journal of Astrophysics and Astronomy, 42, 89

\bibitem[{{Jadhav} {$et~al$.}(2019){Jadhav}, {Sindhu}, \& {Subramaniam}}]{Jadhav2019ApJ...886...13J}
{Jadhav}, V.~V., {Sindhu}, N., \& {Subramaniam}, A. 2019, \apj, 886, 13

\bibitem[{{Jadhav} {$et~al$.}(2023){Jadhav}, {Subramaniam}, \& {Sagar}}]{Jadhav2023A&A...676A..47J}
{Jadhav}, V.~V., {Subramaniam}, A., \& {Sagar}, R. 2023, \aap, 676, A47

\bibitem[{{Koester}(2010)}]{Koester2010MmSAI..81..921K}
{Koester}, D. 2010, \memsai, 81, 921

\bibitem[{{Levenhagen} {$et~al$.}(2017){Levenhagen}, {Diaz}, {Coelho}, \& {Hubeny}}]{Levenhagen2017ApJS..231....1L}
{Levenhagen}, R.~S., {Diaz}, M.~P., {Coelho}, P. R.~T., \& {Hubeny}, I. 2017, \apjs, 231, 1

\bibitem[{{Lodieu} {$et~al$.}(2019){Lodieu}, {Allard}, {Rodrigo}, {Pavlenko}, {Burgasser}, {Lyubchik}, {Kaminsky}, \& {Homeier}}]{Lodieu2019A&A...628A..61L}
{Lodieu}, N., {Allard}, F., {Rodrigo}, C., {$et~al$.} 2019, \aap, 628, A61

\bibitem[{{Luo} {$et~al$.}(2021){Luo}, {Zhao}, {Li}, {Guo}, \& {Liu}}]{Luo2021arXiv210811120L}
{Luo}, F., {Zhao}, Y.-H., {Li}, J., {Guo}, Y.-J., \& {Liu}, C. 2021, Research in Astronomy and Astrophysics, 21, 272

\bibitem[{{Martin} {$et~al$.}(2005){Martin}, {Fanson}, {Schiminovich}, {Morrissey}, {Friedman}, {Barlow}, {Conrow}, {Grange}, {Jelinsky}, {Milliard}, {Siegmund}, {Bianchi}, {Byun}, {Donas}, {Forster}, {Heckman}, {Lee}, {Madore}, {Malina}, {Neff}, {Rich}, {Small}, {Surber}, {Szalay}, {Welsh}, \& {Wyder}}]{Martin2005ApJ...619L...1M}
{Martin}, D.~C., {Fanson}, J., {Schiminovich}, D., {$et~al$.} 2005, \apjl, 619, L1

\bibitem[{{Moe} \& {Di Stefano}(2017)}]{Moe2017ApJS..230...15M}
{Moe}, M., \& {Di Stefano}, R. 2017, \apjs, 230, 15

\bibitem[{{Morley} {$et~al$.}(2014){Morley}, {Marley}, {Fortney}, {Lupu}, {Saumon}, {Greene}, \& {Lodders}}]{Morley2014ApJ...787...78M}
{Morley}, C.~V., {Marley}, M.~S., {Fortney}, J.~J., {$et~al$.} 2014, \apj, 787, 78

\bibitem[{{Nine} {$et~al$.}(2023){Nine}, {Mathieu}, {Gosnell}, \& {Leiner}}]{Nine2023ApJ...944..145N}
{Nine}, A.~C., {Mathieu}, R.~D., {Gosnell}, N.~M., \& {Leiner}, E.~M. 2023, \apj, 944, 145

\bibitem[{{Oliphant}(2007)}]{Oliphant2007CSE.....9c..10O}
{Oliphant}, T.~E. 2007, Computing in Science and Engineering, 9, 10

\bibitem[{{Pacheco} {$et~al$.}(2021){Pacheco}, {Diaz}, {Levenhagen}, \& {Coelho}}]{Pacheco2021ApJS..256...41P}
{Pacheco}, T.~A., {Diaz}, M.~P., {Levenhagen}, R.~S., \& {Coelho}, P. R.~T. 2021, \apjs, 256, 41

\bibitem[{{Pal} {$et~al$.}(2024){Pal}, {Subramaniam}, {Reddy}, \& {Jadhav}}]{Pal2024ApJ...970L..39P}
{Pal}, H., {Subramaniam}, A., {Reddy}, A. B.~S., \& {Jadhav}, V.~V. 2024, \apjl, 970, L39

\bibitem[{{Panthi} {$et~al$.}(2024){Panthi}, {Vaidya}, {Vernekar}, {Subramaniam}, {Jadhav}, \& {Agarwal}}]{Panthi2024MNRAS.527.8325P}
{Panthi}, A., {Vaidya}, K., {Vernekar}, N., {$et~al$.} 2024, \mnras, 527, 8325

\bibitem[{{Perez} \& {Granger}(2007)}]{Perez2007CSE.....9c..21P}
{Perez}, F., \& {Granger}, B.~E. 2007, Computing in Science and Engineering, 9, 21

\bibitem[{{Rao} {$et~al$.}(2022){Rao}, {Vaidya}, {Agarwal}, {Panthi}, {Jadhav}, \& {Subramaniam}}]{Rao2022MNRAS.516.2444R}
{Rao}, K.~K., {Vaidya}, K., {Agarwal}, M., {$et~al$.} 2022, \mnras, 516, 2444

\bibitem[{{Sindhu} {$et~al$.}(2019){Sindhu}, {Subramaniam}, {Jadhav}, {Chatterjee}, {Geller}, {Knigge}, {Leigh}, {Puzia}, {Shara}, \& {Simunovic}}]{Sindhu2019ApJ...882...43S}
{Sindhu}, N., {Subramaniam}, A., {Jadhav}, V.~V., {$et~al$.} 2019, \apj, 882, 43

\bibitem[{{Skrutskie} {$et~al$.}(2006){Skrutskie}, {Cutri}, {Stiening}, {Weinberg}, {Schneider}, {Carpenter}, {Beichman}, {Capps}, {Chester}, {Elias}, {Huchra}, {Liebert}, {Lonsdale}, {Monet}, {Price}, {Seitzer}, {Jarrett}, {Kirkpatrick}, {Gizis}, {Howard}, {Evans}, {Fowler}, {Fullmer}, {Hurt}, {Light}, {Kopan}, {Marsh}, {McCallon}, {Tam}, {Van Dyk}, \& {Wheelock}}]{Skrutskie2006AJ....131.1163S}
{Skrutskie}, M.~F., {Cutri}, R.~M., {Stiening}, R., {$et~al$.} 2006, \aj, 131, 1163

\bibitem[{{Tandon} {$et~al$.}(2017){Tandon}, {Subramaniam}, {Girish}, {Postma}, {Sankarasubramanian}, {Sriram}, {Stalin}, {Mondal}, {Sahu}, {Joseph}, {Hutchings}, {Ghosh}, {Barve}, {George}, {Kamath}, {Kathiravan}, {Kumar}, {Lancelot}, {Leahy}, {Mahesh}, {Mohan}, {Nagabhushana}, {Pati}, {Kameswara Rao}, {Sreedhar}, \& {Sreekumar}}]{Tandon2017AJ....154..128T}
{Tandon}, S.~N., {Subramaniam}, A., {Girish}, V., {$et~al$.} 2017, \aj, 154, 128

\bibitem[{{The pandas development Team}(2024)}]{pandas2022zndo...3509134T}
{The pandas development Team}. 2024, {pandas-dev/pandas: Pandas}, doi:10.5281/zenodo.3509134

\bibitem[{{Thompson} {$et~al$.}(2021){Thompson}, {Frinchaboy}, {Spoo}, \& {Donor}}]{Thompson2021AJ....161..160T}
{Thompson}, B.~A., {Frinchaboy}, P.~M., {Spoo}, T., \& {Donor}, J. 2021, \aj, 161, 160

\bibitem[{{Tremblay} {$et~al$.}(2011){Tremblay}, {Bergeron}, \& {Gianninas}}]{Tremblay2011ApJ...730..128T}
{Tremblay}, P.~E., {Bergeron}, P., \& {Gianninas}, A. 2011, \apj, 730, 128

\bibitem[{{Virtanen} {$et~al$.}(2020){Virtanen}, {Gommers}, {Oliphant}, {Haberland}, {Reddy}, {Cournapeau}, {Burovski}, {Peterson}, {Weckesser}, {Bright}, {van der Walt}, {Brett}, {Wilson}, {Millman}, {Mayorov}, {Nelson}, {Jones}, {Kern}, {Larson}, {Carey}, {Polat}, {Feng}, {Moore}, {VanderPlas}, {Laxalde}, {Perktold}, {Cimrman}, {Henriksen}, {Quintero}, {Harris}, {Archibald}, {Ribeiro}, {Pedregosa}, {van Mulbregt}, \& {SciPy 1. 0 Contributors}}]{Virtanen2020NatMe..17..261V}
{Virtanen}, P., {Gommers}, R., {Oliphant}, T.~E., {$et~al$.} 2020, Nature Methods, 17, 261

\bibitem[{{Wright} {$et~al$.}(2010){Wright}, {Eisenhardt}, {Mainzer}, {Ressler}, {Cutri}, {Jarrett}, {Kirkpatrick}, {Padgett}, {McMillan}, {Skrutskie}, {Stanford}, {Cohen}, {Walker}, {Mather}, {Leisawitz}, {Gautier}, {McLean}, {Benford}, {Lonsdale}, {Blain}, {Mendez}, {Irace}, {Duval}, {Liu}, {Royer}, {Heinrichsen}, {Howard}, {Shannon}, {Kendall}, {Walsh}, {Larsen}, {Cardon}, {Schick}, {Schwalm}, {Abid}, {Fabinsky}, {Naes}, \& {Tsai}}]{Wright2010AJ....140.1868W}
{Wright}, E.~L., {Eisenhardt}, P. R.~M., {Mainzer}, A.~K., {$et~al$.} 2010, \aj, 140, 1868

\end{thebibliography}

\end{document}